\documentclass[prd,aps]{revtex4}
\usepackage{graphicx}
\usepackage{caption}

\begin{document}

%\preprint{}

\title{Inflationary Tensor Perturbations After BICEP}


\author{Jerod Caligiuri and Arthur Kosowsky}
%\thanks{}
\affiliation{Department of Physics and Astronomy, University of Pittsburgh, Pittsburgh, PA 15260}
\affiliation{Pittsburgh Particle Physics, Astrophysics, and Cosmology Center (Pitt-PACC), Pittsburgh, PA 15260}


\date{March 19, 2014}

\begin{abstract}
The measurement of B-mode polarization of the cosmic microwave background at
large angular scales by the BICEP experiment suggests a stochastic gravitational
wave background from early-universe inflation with a surprisingly large amplitude. The
power spectrum of these tensor perturbations can be probed 
both with further measurements of the microwave background polarization at
smaller scales, and also directly via interferometry in space. We show that sufficiently
sensitive high-resolution  B-mode measurements will ultimately have the ability to 
test the inflationary consistency relation between the amplitude and spectrum of the 
tensor perturbations, confirming their inflationary origin. Additionally, a precise B-mode
measurement of the tensor spectrum
will predict the tensor amplitude on solar system scales to 20\% accuracy for an exact 
power law tensor spectrum, so a direct
detection will then measure the running of the tensor spectral index to high precision. 
\end{abstract}


\maketitle


The remarkable observations of the BICEP experiment \cite{BICEP2014}, if correct, may reveal the existence of 
tensor perturbations in the universe with an unexpectedly large amplitude. The measured B-mode component
of the polarization
power spectrum \cite{KKS1997,ZaldSelj1997} is consistent with a scale-invariant gravitational wave background with
a tensor-scalar ratio of $r=0.2$. These tensor perturbations
could arise from inflation in the early universe, but further characterization of the signal is needed to
make this case compelling. The large amplitude of the signal creates a realistic possibility for
two independent tests of an inflationary origin for the tensor perturbations: one, higher precision
measurements of the B-mode polarization, and two, the direct detection of the gravitational
wave background with space-based interferometry. Previous work has considered this pairing
of experiments as an inflation probe \cite{Turner1997,Ungarelli2005,Cooray2005,SmithKamCoor2006,SmithPeirisCoor2006,ChongEfst2006, Friedman2006}, 
but their combination becomes far more informative if the amplitude of the tensor perturbations is as large as $r=0.2$.

Inflation generally produces a power-law power spectrum for both scalar and tensor perturbations, 
$P_S(k)=A_S\left(k/k_0\right)^{n_S-1}$ 
and $P_T(k)=A_T \left(k/k_0\right)^{n_T}$ (see, e.g., the classic review \cite{Lidsey1997}). The scalar perturbation
amplitude $A_S$ and spectral index $n_s$ have been determined to high precision through measurements
of the microwave background temperature anisotropies \cite{Sievers2013,Hou2014,PlanckParams2013}. Thus the amplitude
of tensor perturbations is generally characterized by the tensor-scalar ratio, $r \equiv P_T/P_S$, 
evaluated at the fiducial wavenumber $k_0\equiv 0.002$  Mpc$^{-1}$.

The simplest models of inflation, which involve a single dynamical degree of freedom evolving
slowly compared to the expansion rate of the universe (single field, slow-roll models) 
predict a relation between the tensor-to-scalar ratio and the tensor power law index known as the \emph{consistency relation},
$r=-8n_T$ \cite{Liddle92}.  This connection arises because both the tensor and scalar power spectrum arise from the single degree of freedom.
If tensor perturbations with $r=0.2$ are generated by inflation, the naive expectation is $n_T = -0.025.$

This value for $n_T$ will be observable with anticipated microwave background polarization experiments. Our
ability to measure $n_T$ is limited by cosmic variance, which provides a fundamental limit to how well the tensor power
spectrum can be measured: we only have a single sky to measure. 
The cosmic variance of the B-mode polarization power spectrum multipole $C_l^{BB}$ is approximately
$\sigma_l = \sqrt{2/(2l+1)f_{\rm sky}}C_l^{BB}$, where $f_{\rm sky}$ is the fraction of the full sky mapped by a given
experiment.  In addition to tensor perturbations, gravitational lensing of the larger E-mode polarization component will produce B-mode
polarization contributing to this cosmic variance \cite{ZaldarriagaSeljak1998}. However, with sufficiently sensitive high-resolution polarization maps,
the lensing signal can be measured directly using the characteristic non-gaussian distribution of polarization which
lensing creates. Knox and Song \cite{KnoxSong2002} originally estimated how well the polarization field can be ``delensed'' using
a quadratic maximum-likelihood estimator of Hu and Okamoto \cite{HuOkamoto2001}, finding cosmic variance due
to the residual lensing signal of roughly $10\%$ for a perfect sky map. Subsequent work by Hirata and Seljak \cite{HirataSeljak2003}
demonstrated that an iterative application of the quadratic estimator can push delensing significantly further, given maps of
sufficient sensitivity and angular resolution. The ability of an experiment with very low noise to measure $n_T$ from a B-mode polarization
map will be determined by the cosmic variance from the sum of the primordial tensor signal plus the residual lensing signal after
any delensing procedure. 

Here we estimate the ability of several nominal future polarization experiments to constrain $n_T$ and
test the consistency relation (see also \cite{SongKnox2003,Verde2006,CMBPol2009,Farhang2011} 
for similar estimations of $n_T$ constraints for weaker signals). Sensitivity and
angular resolutions of these experiments are given in Table 1; for all cases we assume a sky coverage of $f_{\rm sky}=0.5$,
corresponding to a single ground-based experiment. We assume no foreground contamination or systematic
errors; to make these assumptions more believable for a ground-based experiment, 
we only consider power spectrum measurements with $l>50$,
corresponding to angular scales smaller than 4 degrees. (A detailed foreground study for future
polarization satellite missions
\cite{Dunkley2008} estimates that, for $r=0.2$, the tensor B-mode signal will dominate over the foreground B-mode signal
for more than 75\% of the sky, so the assumption of foreground-cleaned maps from 
multi-frequency experiments reasonable.) 
We also assume the theoretical power spectrum of
gravitational lensing is known exactly, which will be a good assumption for upcoming experiments based on our
knowledge of structure growth in the standard cosmological model. Uncertainties in the lensing model 
(currently around $2\%$ in the lensing amplitude, e.g.\ Fig.~12 of \cite{BICEP2014}) will only cause small changes to these results.
We assume the residual lensing signal amplitude given by Seljak and Hirata \cite{HirataSeljak2003} (listed in Table 1) 
for the given map sensitivities and angular resolution.

\begin{table} [h!]
\begin{center} 
\begin{tabular}{c|c|c|c} 
\hline 
Experiment & $\sigma_{Q,U}$ ($\mu$K-arcmin) & beam (FWHM) (arcmin) & lensing residual \cite{HirataSeljak2003}\\ 
\hline 
Example A & $1.41$ &  $4$ & $10\%$ \\ 
\hline 
Example B & $0.5$ & $4$ & $5\%$ \\ 
\hline 
Example C & $0.25$ &  $4$ & $1\%$ \\
\hline
\end{tabular} 
\caption{Parameters of model polarization experiments.} 
\label{Tab1} 
\end{center}
\end{table}

Figure 1 displays allowed values for each experiment in the $r$-$n_T$ plane, for a fiducial model with $r=0.2$
and $n_T= -0.025$ satisfying the inflationary consistency relation. These have been calculated using
a simple quadratic likelihood evaluated on a grid of models in the parameter plane. We compute $C_l^{BB}$
using the CAMB package \cite{CAMB1} and use only multipoles $50 < l < 2000$ 
in computing the likelihoods, with a pivot scale $k_0=0.002$  Mpc$^{-1}$ . 
Including higher multipoles has a negligible effect on the likelihoods; the 
inclusion of lower multipoles also has only a small effect, due to cosmic variance. Improving angular 
resolution to 2 arcminutes will decrease the lensing residual marginally, by roughly $10\%$ (see Fig.~5 in
Ref. \cite{HirataSeljak2003}). Constraints on $n_T$ improve
dramatically as the map sensitivity increases from 1.4 $\mu$K-arcmin to 0.25 $\mu$K-arcmin, due to better
delensing. At the
lower sensitivity value, given by Example C, $n_T$ has a normal error of around $0.006$, so
is measured to be different from $0$ at around $4\sigma$. 
With perfect cleaning of lensing, the significance away from $0$ for Example C would increases marginally
from $4\sigma$ to $5\sigma$, but improving
on the residual lensing contribution in Example C requires a more sophisticated treatment of delensing \cite{HirataSeljak2003}. 
The same sensitivity and resolution for a full-sky map, presumably from a satellite mission, 
would increase sky coverage by a factor of 2,
decreasing the cosmic-variance limited errors by a factor of $\sqrt{2}$, and give a normal error on $n_T$
of around 0.004, constraining $n_T$ away from zero at a $6\sigma$ level. 

If the actual amplitude of the tensor perturbations turns out to be $r=0.1$ instead of $r=0.2$ (consistent with
an alternate foreground dust model in \cite{BICEP2014}), then the
consistency-relation value of $n_T$ decreases by a factor of 2, while the B-mode signal used to measure $n_T$
also decreases in amplitude by a factor of 2. Then a full-sky map with the sensitivity of Example C 
provides a determination of $n_T$ with error around $0.006$, which is now different from 0 at a $2\sigma$ significance. 
The consistency relation still passes a strong test, but we are not able to distinguish between a consistency-relation 
inflation model and a naive scale-invariant tensor background of unspecified origin. 

\begin{figure}[]
\includegraphics[scale=0.6,angle=180]{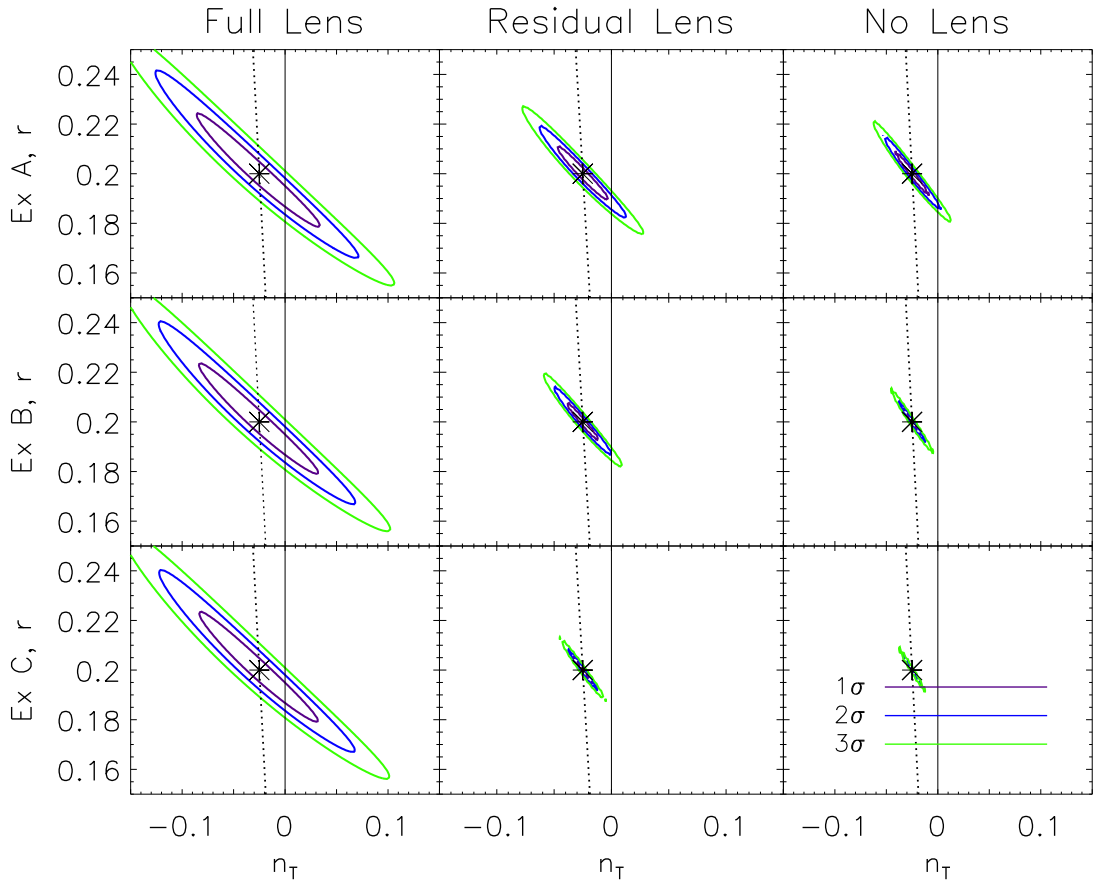}
\caption{Likelihood countours in the $r$-$n_T$ plane for the model polarization experiments in Table 1. The fiducial
model is $r = 0.2$ and $n_T = -0.025$, indicated by $*$. The dotted line indicates the inflation consistency relation.  
The vertical line indicates $n_T=0$ for reference. (Left) Full lensing contribution to the cosmic variance error. (Center) Residual lensing contribution after
delensing (see Table 1 for residual lensing noise levels), and (Right) No lensing contribution to the cosmic variance, for comparison.}
\label{FIG1}
\end{figure}

The example experiments in Table 1 represent a range encompassing possible sensitivities for a so-called
``Stage 4'' microwave background experiment \cite{DETF2006,Snowmass2013}. The ability to measure $n_T$ depends strongly
on the sensitivity in this range. Polarization maps with 4 arcminute resolution or better and map sensitivity
of 0.25 $\mu$K-arcmin or better can decisively test the inflationary consistency relation between $r$ and $n_T$
for the BICEP value of $r=0.2$. A measurement of $n_T$ obeying the consistency relation and inconsistent
with the generic scale-invariant spectrum $n_T=0$ would provide highly non-trivial evidence in
favor of the tensor perturbations arising from a simple single-field, slow-roll inflation epoch in the early universe.
Note that a value of $n_T$ different from the inflation consistency relation would not necessarily rule out inflation as the
source of the tensor perturbations, but could alternately give valuable information that the inflation mechanism
was more complicated than a simple slow-roll model, provided the tensor perturbations did arise from inflation.

The second test of inflation is a direct detection of the tensor perturbations using space-based
interferometry. A stochastic background of gravitational waves with a scale-invariant power-law
spectrum and an amplitude of $r=0.2$ results in a tensor energy density of $\Omega_{\rm GW}h^2\simeq 10^{-15}$. 
Fortuitously, this tensor amplitude is about the value which
maximizes the direct detection amplitude at terrestrial scales \cite{Turner1997,Maggiore2000}. 
NASA's Big Bang Observer (BBO) concept study \cite{BBOconcept}, 
an extension of the Laser Interferometer Space Antenna (LISA) 
proposal to higher sensitivities and shorter satellite separations, would detect this
signal at a frequency of $0.1$ Hz with a significance of $100\sigma$ with one year of observing \cite{Kudoh2006,Seto2006}, in the
absence of confusion noise from white-dwarf binaries at cosmological distances \cite{Seto2006}. Such a binary background
will be isotropic with a steeply falling power spectrum; a nonisotropic binary contribution from the Milky
Way would also need to be accounted for. A second more speculative
stochastic gravitational wave contribution could arise from a possible early-universe phase transition
at the electroweak scale \cite{KamKosTurn1994,Gogo2007}. For high-precision characterization
of the primordial signal, the competing backgrounds would either need to be modeled with
comparable precision, or interferometric measurements would need to have a low-frequency cutoff, 
reducing the detection significance for BBO substantially \cite{Kudoh2006}. 
Experiments with greater sensitivity and higher frequency ranges than BBO have been contemplated:
Kudoh et al. \cite{Kudoh2006} have calculated that with an 0.2 Hz lower frequency
cutoff, their ``Fabry-Perot DECIGO'' \cite{DECIGO2010,DECIGO2011} would detect the $r=0.2$ 
gravitational wave background at $10\sigma$, while
their ``Ultimate DECIGO''  \cite{DECIGO} would detect it at $5\times 10^4\sigma$. 
%Clearly a $100\sigma$ detection
%of the tensor mode signal is not out of the question. 
%There appears to be no fundamental physical
%limit to the sensitivity of direct interferometric gravitational wave measurements \cite{XX}.

The B-mode polarization of the microwave background arises from tensor modes with a 
characteristic wavelength of $k_0^{-1}\simeq 100$ Mpc,
while direct detection experiments probe characteristic wavelengths of $c/\nu = 2\times 10^{-2}$ A.U., 
a range covering a factor of $10^{15}$ in wavelength.  A determination of $r$ and $n_T$ in the
B-mode power spectrum means that the tensor spectrum can be extrapolated to smaller
wavelengths, assuming a perfect power law spectrum. 
The amplitude at a smaller scale will have an uncertainty governed by
the uncertainties in $r$ and $n_T$ at the larger scale, like those displayed in Fig.~1. If $r=0.2$, 
a full-sky B-polarization
map with sensitivity below 1 $\mu$K arcmin will determine $n_T= -0.025$ with an error of around
$0.004$. Then extrapolating to a scale $10^{15}$ times smaller in 
wavelength using two different values of $n_T$ differing by 0.004 gives an
amplitude difference of $20\%$. This is much larger than the difference due to uncertainty in $r$
which we can ignore. 

A direct detection experiment which could measure the tensor amplitude to significantly better
than 20\% could thus detect a difference from the predicted amplitude with an uncertainty
of around 20\% of the measured amplitude. Such a difference would arise if the spectrum
is not a perfect power law, but rather has some variation in its power law with scale. In analogy
with the running of the density perturbation spectrum \cite{KosowskyTurn1995}, define the
running of the tensor spectral index $\alpha_T \equiv dn_T/d\ln k$. A value of $\alpha_T = 2\times 10^{-4}$
will result in a difference in amplitude of $20\%$ when extrapolated over a factor of $10^{16}$
in wavelength; so by comparing with the values of
$r$ and $n_T$ measured from B-mode polarization,
a direct measurement of the tensor amplitude with interferometry can
measure $\alpha_T$ with an error of around $2\times10^{-4}$.
(If $r=0.1$, then the error on running becomes weaker by about factor of 2.) Discussions of
tensor perturbations until now have commonly claimed that any measurement
of $\alpha_T$ is hopeless -- but we see that it is likely possible to measure the tensor running
{\it better} than scalar running, {\it if} precise measurements of both
$r$ and $n_T$ are obtained from B-mode polarization. 

Direct detection of the gravitational wave background
with a very high significance, by some future experiment like Ultimate DECIGO,
would allow a second measurement of $n_T$, at a wavelength of around 0.02 A.U. Then
this value could be compared with the spectrum extrapolated from the B-mode $n_T$ plus
the running required to give the measured amplitude at A.U. scales. Consistency 
would demonstrate that a constant-running approximation to the tensor power spectrum is valid,
and further verify the inflationary origin of this background; in this case, we can hope to have
six measured quantities characterizing the physics of inflation (the amplitude, power law index, and running for
both scalar and tensor perturbations). If the two values do not match, it would show
that further parameters aside from a spectral index and a constant running are required to adequately
describe the tensor power spectrum over these scales. 

Inflation was invented to solve a well-known litany of cosmological conundrums: the flatness and isotropy
of the universe and the absence of magnetic monopoles. Inflation also provides a simple mechanism for generating
a primordial gaussian random distribution of nearly scale-invariant density perturbations and phase-coherent acoustic
oscillations, which gives an excellent match to observed microwave background temperature anisotropies and
the large-scale distribution of galaxies. Aside from these well-observed signals, inflation
makes a completely generic prediction of a nearly scale-invariant background of tensor perturbations, which
once generated will propagate unimpeded  until the present day, and unchanged save for their dilution and stretching due to
the expansion of the universe. But the amplitude of this stochastic background depends
on the unknown energy scale of inflation. Prior to St.~Patrick's Day 2014, we had only suggestive and somewhat
controversial theoretical reasons that the tensor amplitude was large enough ever to detect. If the 
B-mode polarization measured by BICEP is due to inflationary tensor perturbations, their amplitude
is a factor of 10 to 100 larger than cautious optimists had hoped. As a result, we suddenly and unexpectedly have
within our reach the chance to probe the unknown physics governing the universe at an age
of $10^{-36}$ seconds and an energy scale of $10^{16}$ GeV, with two completely different experimental approaches. 
We are surely obliged to try. 

\acknowledgments
We thank Jeff Newman for helpful conversations on statistics, Simone Aiola for helpful discussions of power 
spectra and aid in using the CAMB package, and James Zibin for pointing out a numerical error in an earlier version of this paper.
We acknowledge the use of the Legacy Archive for Microwave Background Data Analysis (LAMBDA), 
part of the High Energy Astrophysics Science Archive Center (HEASARC). 
HEASARC/LAMBDA is a service of the Astrophysics Science Division at the NASA Goddard Space Flight Center.  
This work has also used NASA's Astrophysical Data System for bibliographic information. The authors are
supported by NSF grant 1312380 through the Astrophysics Theory Program.

\bibliography{sources}


\end{document}